\documentclass{sig-alternate-10pt}
\usepackage{enumerate}
\usepackage[utf8]{inputenc}
\usepackage{amsmath}
\usepackage{amsfonts,dsfont}
\usepackage{amssymb}
\usepackage[hyphens]{url}
\usepackage{hyperref}
\usepackage{graphicx}
\usepackage{subfig}
\usepackage{color}
\usepackage{epstopdf}
\usepackage{multirow}
\usepackage{listings}
\usepackage{clrscode}
\usepackage{xcolor}
\usepackage[boxed]{algorithm}
\usepackage{algorithmic}
\usepackage[normalem]{ulem}
\usepackage{comment}
\DeclareCaptionType{copyrightbox}

\newcommand{\KDAF}{\emph{PKA}}

\newcommand{\Protect}{\emph{PROTECT}}
\newcommand{\Access}{\emph{ACCESS}}
\newcommand{\V}{\mathcal{V}}
\newcommand{\Attr}{\mathcal{W}} 
\newcommand{\n}{n}
\newcommand{\threshold}{t}
\newcommand{\M}{\mathcal{P}} 
\newcommand{\PM}{\mathcal{C}} 
\newcommand{\key}{K}
\newcommand{\enc}{Enc}
\newcommand{\dec}{Dec}
\newcommand{\SSS}{SSS}
\newcommand{\distr}{\mathds{D}}
\newcommand{\SSScreate}{\emph{SSS-CREATE}}
\newcommand{\SSSreconstruct}{\emph{SSS-RECONSTRUCT}}
\newcommand{\K}{\mathcal{K}}
\newcommand{\C}{C}
\newcommand{\kap}{\kappa}

\newcommand{\pr}[1]{\mathrm{Pr}[#1]}

\newcommand{\gf}{$GF(2^8)$}
\newcommand{\adv}{\mathcal{A}} 
\newcommand{\ora}{\mathcal{O}} 
\newcommand{\A}[3]{$\mathcal{A}^{\distr_B(\mathrm{#1})}_#2(\distr_T(\mathrm{#3}))$}
\newcommand{\Aplain}[3]{$\mathcal{A}^{#1}_#2(#3)$}

\newcommand{\fnsize}[1]{\scriptsize{#1}}
\newcommand{\alsize}[1]{\small{#1}}

\definecolor{DarkBlue}{rgb}{0.1,0.1,0.7}
\newcommand{\schange}[2]{#2}

\newtheorem{theorem}{Theorem}[section]

\newtheorem{definition}[theorem]{Definition}
\newtheorem{remark}[theorem]{Remark}

\title{Protecting Public OSN Posts from Unintended Access}

\numberofauthors{2}
\author{
Frederik Armknecht and Manuel Hauptmann\\
      \affaddr{Universit\"at Mannheim}\\
      \email{\normalsize{\{armknecht|mhauptma\}@uni-mannheim.de}}
\and
\alignauthor 
   Stefanie Roos and Thorsten Strufe\\
      \affaddr{TU Darmstadt \& CASED}\\
      \email{\normalsize{\{firstname.lastname\}@cased.de}}
}

\begin{document}
\maketitle

\begin{abstract}

The design of secure and usable access schemes to personal data represent a major challenge of online social networks (OSNs). State of the art requires prior interaction to grant access.  Sharing with users who are not subscribed or previously have not been accepted as contacts in any case is only possible via public posts, which can easily be abused by automatic harvesting for user profiling, targeted spear-phishing, or spamming. Moreover, users are restricted to the access rules defined by the provider, which may be overly restrictive, cumbersome to define, or insufficiently fine-grained.

We suggest a complementary approach that can be easily deployed in addition to existing access control schemes, does not require any interaction, and includes even public, unsubscribed users. It exploits the fact that different social circles of a user share different experiences and hence encrypts arbitrary posts. Hence arbitrary posts are encrypted, such that only users with sufficient knowledge about the ow\-ner can decrypt. 

Assembling only well-established cryptographic primitives, we prove that the security of our scheme is determined by the entropy of the required knowledge. We consequently analyze the efficiency of an informed dictionary attack and assess the entropy to be on par with common passwords. A fully functional implementation is used for performance evaluations, and available for download on the Web.

\end{abstract}
\section{Introduction}
\label{sec:intro}

Profiles and posts in Online Social Networks (OSN)  reveal a wealth of personal information about their owners \cite{quercia11privacy}. This is a common, and very immediate threat as users still casually lose control over personal data like addresses and even the state of their health.
For instance, compromising photos or personal opinions have been abused for commercial \footnote{\fnsize{\url{http://www.iwf.org.uk/about-iwf/news/post/334-young-people-are-warned-they-may-lose-control-over-their-images-and-videos-once-they-are-uploaded-online}}} and criminal use \footnote{\fnsize{\url{http://www.telegraph.co.uk/technology/news/8789538/Most-burglars-using-Facebook-and-Twitter-to-target-victims-survey-suggests.html}}}.
%
Furthermore, automatic harvesting of profiles and posts by Web servers in data centers yields large scale automatic access: directly at the database by the provider and institutional parties, through API calls by their affiliates, or through simple Web crawlers of arbitrary, external parties. 
Such mass retrieval amplifies the threats to mass mailing-, phishing-, and scamming campaigns at very large 
scales\footnote{\fnsize{\url{http://www.theregister.co.uk/2012/10/31/dodgy_brokering_facebook_data/}}}.

The most common approach for ensuring explicit audience selection is the use of access control mechanisms by configuring provider-defined access rules. 
%
However, this automatically denies access to non-members and user who have not interacted within the OSN previously.
In consequence, granting access to individuals from the greater social community other than those that explicitly are declared as friends within the OSN, or the intransparent and uncontrolled set of ``friends-of-friends'', is solely possible by broadcasting a post without any access restriction to the broad public.
But such public posts impose the enormous risk of access by entirely unknown and unintended parties, predominantly the afore-mentioned crawlers. 
Hence, sharing exclusively with the extended social community of a user is not sufficiently supported today.


\subsection{Contribution}
In this paper, we close this gap by presenting {\em Partial Knowledge-based Access Control (PKA)}, a non-inter\-active access control scheme that can be implemented in addition to existing mechanisms.  
Its main purpose is to protect  public posts from unintended access by unknown parties.
No out-of-band key exchange is necessary, allowing even friends that are not subscribed to the OSN to access the posts.

The main idea is to encrypt the post using attribute values that are easy to guess for members of a user's social community, yet difficult for strangers and during automated access. As in general even a member of this community may not have knowledge of all values, such a requirement would be too strict in most cases. Instead our scheme provides a higher level of flexibility: 
For each post, the publisher can \emph{individually} choose $\n$ attributes as well as
the number of attributes a user has to know to access the posts. 

We describe a concrete instantiation based on established cryptographic mechanisms. 
We prove that if these mechansims are secure, the security of \KDAF\ is completely determined by  the entropy of the chosen attributes. 

Based on data, we acquired from well-known OSNs and the German statistical office,
we find that, even under an advanced and informed dictionary attack, on the order of hundreds of millions of potential attribute combinations have to be tested on average to access profiles that are protected with \KDAF.
This is about the same order of effort as it is needed to break passwords \cite{malone:password,Narayanan05fast}.

Hence, our scheme  protects against common attackers such as scammers and spammers, as well as against
casual data loss, and leaks by the provider and governmental agencies. 
Observe that, depending on the particular use case, the owner may choose attributes of higher entropy, 
which are harder or virtually impossible to guess for anybody.
In fact, the traditional approach of encrypting a post using a shared secret key can be seen as a special case of our approach.

We provide a fully functional Firefox extension for Facebook posts as a prototype, for public download. 
Locally storing the attribute-value pairs that are used as secrets once they are provided, the profiles containing protected posts can be accessed repeatedly without the need for repeated entry.
The overhead of the cryptographic scheme is barely perceived by a legitimate user,
yet delays unauthorized access by several days, rendering automatic harvesting by
both external parties and social network providers economically unviable. 

Section \ref{sec:relatedwork} covers the state of the art. We describe the general framework of our approach and give a formal security definition in Section \ref{sec:concept}. Afterwards, we explain a concrete realization in Section \ref{sec:protocol}. Its security and analysis are analyzed in Sections \ref{sec:analysis} and \ref{sec:plugin}, respectively, and Section \ref{sec:conclusion} concludes the paper.

\section{Related Work}
\label{sec:relatedwork}

OSN privacy is constantly raising broad interest, and recent years have seen various contributions aiming at preserving it.
The approaches differ by their assumed trust model and the adversaries they aim to defend against.

Decentralized social networks, like for instance {\em Vis-a-Vis} \cite{shakimov09privacy}, {\em diaspora}
\footnote{\fnsize{\url{http://joindiaspora.com}}}, 
and {\em Safebook} \cite{cutillo09safebook} have been proposed to circumvent the existence of a centralized instance that has a global overview of all subscribers, as well as ubiquitous control.
These systems require migration from current services, additionally, their lack of acceptance, partially caused by their lack of competitive functionality and performance, have prevented broad application, so far. 

Focusing on confidentiality of content rather than anonymity or unobservability, and thus accepting that acts of communication as well as their participators can be observed by the social network provider, a second class of approaches encrypts content and leverages the conventional service for its exchange.
{\em NOYB} \cite{Guha08noyb} as one of the earliest approaches introduces dictionaries to replace the real with seemingly random attributes. 
The correct mapping is shared with authorized contacts, who hence are capable of retrieving the plaintext.
This mapping can be retrieved by the social networking provider, too, though, when acting on behalf of a user's friend.
{\em Persona} \cite{Baden09persona} and {\em EASiER} \cite{Jahid2011easier} employ attribute-based encryption and offer fine grained, personalized access rules to different parts of the profile. The Firefox extension Scramble! \cite{beato2011scrample} offers item-specific access rules over multiple social networking sites, but again requires the user
to explicitly define friendship relations in several social networks. 

Facebook indeed has started to prevent sharing of encrypted content by transcoding image files and restricting free-text attribute fields.
Some recent approaches hence suggest the exchange of encrypted content via third party storage, using the OSN only for exchange of references to the encrypted content \cite{lucasB09,tootoonchian09lockr,luo09facecloak,fahl12trustsplit}. Alternatively, the transmission of encrypted content can be hidden \cite{ion13someeyes}.
All these approaches require interaction for explicit authorization on a per-user base and key exchange.

Approaches that aim at preventing unnoticed access by unintended parties again differ in the adversary they assume.
PoX \cite{egele12pox} is a plugin that illustrates which information is provided to third-party applications. 
It additionally offers the possibility to restrict access of these parties to selected profile entries.  
Our goal differs from theirs: they do not actually want to hide content from general, unintended audiences, but rather enable the user to consciously decide which data is accessed by application providers.
{\em CAPTCHAs} indeed serve the purpose of preventing automated mass retrieval, but they allow access to arbitrary human individuals that can solve the challenge, thus not only to trusted parties. In a similar fashion, text in OSNs can be distorted to protect against automatic processing \cite{pashalidis12humaneyes}.
Using personal knowledge of and about a user to complement keys and passphrases is leveraged by {\em social authentication}%
\footnote{\fnsize{\url{http://www.facebook.com/blog/blog.php?post=486790652130}}}
to prevent illegitimate login to accounts.
Both CAPTCHAs and social authentication are aimed at adversaries external to the service, though, and thus fail to prevent unintended access by the provider and its affiliates.

In summary, there are various protection schemes that can be used to enhance privacy in OSNs.
In general, they require explicit audience selection, which is desired in case of highly sensitive data.
However, these access rules exclude non-members and may either
be overly restrictive or entail over-sharing due to the heterogeneous contacts summarized
under the common term friends or friends-of-friends.
\section{Framework}
\label{sec:concept}

In this section, we first give a short, informal overview of our requirements
for a knowledge-based access scheme for (public) posts.
Afterwards, the required functionalities are formalized, as are our security goals.

\subsection{Intended Use}

\schange{}{The main goal of our scheme is to protect public post from automatic
harvesting. 
Furthermore, it can also be used to obfuscate information from the social network
provider, its affiliates, and curious strangers.
However, these might have additional knowledge about a user, so that our security analysis
in Section \ref{sec:securityC} does not directly apply.}
The posts of a user should be readable by any person sharing enough experience with the
respective user, regardless of prior interaction within the social network.
Because even non-members can access the posts, but not arbitrary strangers, the peer pressure to join a social network for retrieving important information
is reduced. 
Furthermore,  it is not necessarily to enlist all people that a user wants to share information with as
friends, or sort them into categories, potentially risking confrontation in case of a perceived
misclassification. 
Information can simply be posted as public according to the provided access rules, and be encrypted
with our scheme to prevent strangers, especially crawlers for user profiling and spamming,
from accessing the data.

Because our scheme is intended to overcome the problems of cumbersome explicit audience selection,
transparency and revocation are not considered. 
If desired, they can be provided by 
additional access rules. 
\schange{For example, a post can be published within the friends or members group, and a user can
be informed of any read action.}
{For example, a post can be published within the friends or members group, so that it simply
restricts the access to those contacts within the OSN, who indeed know the publisher well.}

\subsection{Functionality}

On a high level, \KDAF\ allows to transform a post $p$ into a protected post $c$ based on  a set of attributes $(a_1,\ldots,a_\n)$, where each attribute $a_i=(d_i,v_i)$ is composed of  an \emph{attribute description} $d_i$ and an \emph{attribute value} $v_i$. For example, an attribute description could be 'Name' and an attribute value 'John'. In particular two different attributes $a\neq a'$ may share the same attribute description, e.g., 'Name', but have different values 'John' and 'Jane'. 
Vice versa, two distinct attributes can have different descriptions, but the same value.
A protected post $c$ is published together with the used attribute descriptions $d_1,\ldots,d_\n$, and the parameters $\n$ and $\threshold$. 
 Only users who know at least $\threshold$ out of these $\n$ attribute values are able to recover the original post $p$ from the protected post $c$. 
 
 For the sake of simplicity, we assume in the following that the values $\threshold$ and $\n$ with $\threshold\leq \n$ are fixed and publicly known. Let $\M$ be the set of posts,  $\Attr$ be the set of attributes $a$, $\V$ the set of attribute values $v$, and $\PM$ be the set of all protected posts $c$.
 
 The main components of \KDAF{} are the two mechanisms \Protect{}  and \Access{}, which
 allow users to protect their posts and access the
 posts of another known user, respectively. 
 They are formally defined
 in Algorithm \ref{algo:kdaf}.
 \Protect{} takes a post $p \in \M$ and $\n$ values for generating
 the protected post $c$.
 The operation \Access{}  maps
 $\threshold$ values and the protected post $c$ to a post $\tilde{p}$.
 In case that at least $\threshold$ values correctly correspond to the values used
 for protecting post $p$, $\tilde{p}=p$ is correctly retrieved, and a random post $\tilde{p}$ is returned otherwise.
 
 \begin{algorithm}
 \begin{algorithmic}
 \alsize{
 \STATE \textbf{\Protect{}}: $\V^\n \times \M \rightarrow \PM$, $((v_1,\ldots,v_\n),p) \mapsto c $
 }
 \end{algorithmic}
 \begin{algorithmic}
 \alsize{
 \STATE \textbf{\Access}: $\V^\threshold \times \{1,\ldots ,n\}^\threshold \times \PM \rightarrow \M$,
 \STATE \hspace{5em}   $((\tilde{v}_{i_1},\ldots,\tilde{v}_{i_\threshold}),(i_1,\ldots , i_t), c) \mapsto \tilde{p} $
 \STATE \hspace{5em} where $c = \Protect((v_1,\ldots,v_\n),p)$
 \STATE \hspace{5em} and $i_1,\ldots,i_\threshold$ distinct
 \STATE \hspace{5em} $\tilde{p} = \begin{cases} p, \text{ if }  \tilde{v}_{i_j} = v_{i_j} \text{ for } j=1\ldots\threshold \\
                                 \text{a random } r \in \M, \text{ otherwise} \end{cases}$  
  } 
 \end{algorithmic}
 \caption{Basic Functionalities}
 \label{algo:kdaf}
 \end{algorithm}
 
\subsection{Security Goals}
 
Next, we formalize the notion of security for \KDAF. As confidentiality of the post content is the primary security goal, we adopt an established security notion for symmetric encryption schemes: \emph{indistinguishability under chosen-plaintext attack} (cf. \cite{Bellare1997}). This is modelled by the following game involving an attacker $\adv$ and an hypothetical entity named oracle $\ora$:   
\begin{description}
\item[Step 1:] $\ora$ samples an $\n$-tuple $(a_1,\ldots,a_\n)\in\Attr$ with $a_i=(d_i,v_i)$ according to some distribution $\distr$ and hands the attribute  descriptions $(d_1,\ldots,d_\n)$ to  $\adv$. 
\item[Step 2:] $\adv$ generates two different posts $p_0\neq p_1$ (of same length) and gives these to $\ora$. 
\item[Step 3:] $\ora$ flips a random coin $b\in\{0,1\}$, returns $c=\Protect((a_i,v_i)_{i=1}^\n,p_b)$ to $\adv$.
\item[Step 4:] $\adv$ outputs a bit $b^*\in\{0,1\}$.  
\end{description}
The attacker $\adv$ wins the game if $b=b^*$. W.l.o.g., we can assume that $\pr{b=b^*}\geq 1/2$.\footnote{If this is not the case we consider a modified adversary $\overline{\adv}$ who simply invokes $\adv$ and returns the complement $\overline{b^*}$ of the output of $\adv$.} Observe that a simple attack is to pick a random bit for $b^*$, independent of $c$, yielding a winning chance of $1/2$. Hence the \emph{advantage} $\mathrm{Adv}_\adv$ of $\adv$ is defined by how much the success probability deviates from $1/2$:
\begin{equation}
\mathrm{Adv}_\adv:=\pr{b=b^*}-1/2.
\end{equation}
Here the probability is taken over all random coins of $\ora$ and $\adv$. 
Security is then characterized by the maximum advantage taken over all adversaries:
\begin{definition}[Security]\label{def:security}
A \KDAF-scheme is \\
\emph{$(\tau,\varepsilon)$-secure} if it holds for all adversaries $\adv$  with a runtime $\leq \tau$ that $\mathrm{Adv}_\adv\leq \varepsilon$.
\end{definition}
\begin{remark}
Observe that in the security game, we allow an attacker to see the chosen attribute descriptions \emph{before} it has to choose the two posts $p_0\neq p_1$. This is in compliance with the common approach to make an adversary as strong as possible. On the other hand, we disallow to influence the choice of the attributes. This has several reasons. First, the attributes can be seen as the analogon to the encryption key in symmetric encryption schemes. Second, the security definition should also cover cases where the distribution $\distr$ is not (or only partially known) to $\adv$. Third, in case that $\distr$ has a low entropy, this is reflected accordingly in the success probability of $\adv$. 
\end{remark}

\begin{figure}
\centering
\includegraphics[width=\linewidth]{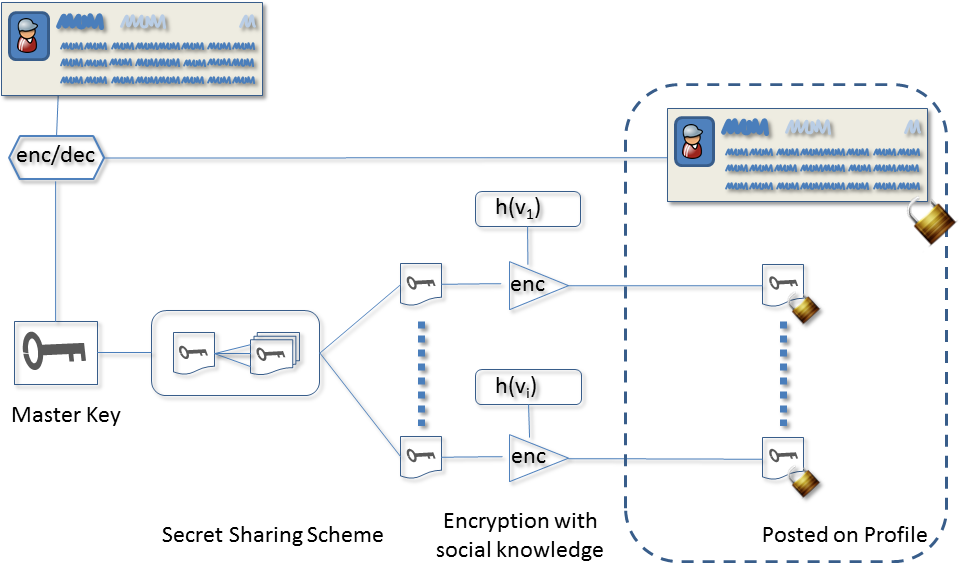}
\caption{Overview of \KDAF{} as described in Section \ref{sec:protocol}}
\label{fig:scheme}
\end{figure}
\section{A Concrete Realization}
\label{sec:protocol}
In this section, we propose a concrete realization of the \KDAF{}-concept 
suggested in Section \ref{sec:concept}.
Our scheme combines established cryptographic primitives: 
\begin{itemize}
\item a block cipher $(\enc,\dec)$ with key space $\K=\{0,1\}^\kap$, which encrypts messages (i.e. posts) from $\M =\{0,1 \}^b$ to ciphertexts in $\C = \M$,\footnote{Remark that fixing the length $b$ of the posts is only done to keep the following description simple. If longer posts need to be encrypted, one can easily adapt the scheme to use the block cipher in an appropriate mode of operation.}
\item a hash function $H:\{0,1\}^* \rightarrow \K$, and 
\item a a secret sharing scheme \SSS{}, consisting of two functions: 
\SSScreate\ creates shares $s_1,\ldots,s_\n\in\K$ out of a secret
$S\in\K$, and $S$ is reconstructed from $\threshold$ shares by \SSSreconstruct{}.
\end{itemize}
In a nutshell, the block cipher is used to encrypt the post using a randomly sampled key $\key$. Using the secret sharing scheme, the key $\key$ is split into $\n$ shares such that any $\threshold$ shares are sufficient for reconstructing $\key$. Each share gets encrypted (using the same block cipher) under individual keys, which are derived from the attribute values $v_i$ and the hash function. 

More formally, during the preparation phase attributes $a_1,\ldots ,a_n$ as well as  integers $\n$ and $\threshold$ are chosen. We assume them to be publicly known, e.g. by attaching them to the protected post.

When a user executes $\Protect{((v_1,\ldots,v_\n),p)}$ for protecting a post $p$, 
a master key $\key$  is uniformly sampled and  $p$ is encrypted to $c=\enc(\key,p)$.
Using the secret sharing scheme, $\n$ shares $(s_1,\ldots,s_\n)$ subsequently are generated from $\key$, and the shares $s_i$ are encrypted to $c_i=\enc(H(v_i),s_i)$ based
on the hashes $H(v_i)$ of the given attribute values $v_i$. The encrypted post $c$ is published together with the encrypted shares $c_i$. 
%
%

When executing the function \Access{} to retrieve the content of an encrypted post $c$, users
provide values $\tilde{v}_{i_j}$ for $\threshold$ of the $\n$ attributes. 
Based on these values, the encrypted shares are then retrieved as follows: for each index $i_j$, the value $H(\tilde{v}_{i_j})$ is computed and afterwards the share $\tilde{s}_{i_j}:=\dec(H(\tilde{v}_{i_j}),s_{i_j})$ is decrypted.
%
A potential key $\tilde{\key}$ is obtained from the shares $\tilde{s}_{i_j}$ by SSSreconstruct{} and the encrypted post is decrypted, i.e. $\tilde{p}=\dec(\tilde{\key},c)$.
If the values $\tilde{v}_{i_j}$ have all been correct, then the correct shares $s_{i_j}$ have been determined, the value $\tilde{\key}$ is equal to $\key$ and in particular $\tilde{p}=p$. Otherwise, that is if at least one of the values $\tilde{v}_{i_j}$ have been wrong, the corresponding value $\tilde{s}_{i_j}$ is wrong with high probability (if the hash function is secure). Due to the properties of the secret sharing scheme, a wrong value $\tilde{K}$ is computed and $\tilde{p}$ will be just a random $b$-bit string. Both algorithms are displayed in Alg.~\ref{algo:concrete}.
A formal security analysis is given in the next Section.

\begin{algorithm}[t]
\begin{algorithmic}
\alsize{
\STATE \textbf{\Protect{}} 
\STATE \textbf{INPUT:} $(v_1,...,v_\n) \in \V^\n$, $p \in \M$
\STATE \hspace{1em} Sample $\key \in \K$
\STATE \hspace{1em} Encrypt $c = \enc(\key,m)$
\STATE \hspace{1em} $(s_1,...,s_\n) = \SSScreate(\key)$
\STATE \hspace{1em} Compute $\key_i = H(v_i)$ for $i=1..n$ 
\STATE \hspace{1em} Encrypt $c_i = \enc(\key_i,s_i)$ for $i=1..n$
\STATE \textbf{OUTPUT:} $(c,c_1,...,c_\n) \in \PM$  
} 
\end{algorithmic}
\vspace{1em}
\begin{algorithmic}
\alsize{
\STATE \textbf{\Access}: 
\STATE \textbf{INPUT:} $(\tilde{v}_{i_1},\ldots,\tilde{v}_{i_\threshold}) \in \V^\threshold$, $(i_1,\ldots , i_{\threshold})$,
\STATE \hspace{4.5em}  $cp = (c,...,c_n)=\Protect{}((v_1,...,v_\n),p)$
\STATE \hspace{1em} Compute $\tilde{\key}_{i_j} = H(\tilde{v}_{i_j})$ for $j=1..\threshold$  
\STATE \hspace{1em} Decrypt $\tilde{s}_{i_j} = \dec(\tilde{\key_{i_j}},c_{i_j})$ for $j=1..\threshold$                                    
\STATE \hspace{1em} $\tilde{\key} = \SSSreconstruct(\tilde{s}_{i_1},...\tilde{s}_{i_\threshold})$
\STATE \hspace{1em} $\tilde{p} = Dec(\tilde{\key},c) = \begin{cases} p, \text{ if }  \tilde{v}_{i_j} = v_{i_j} \text{ for } j=1..\threshold \\
                                \text{a random } r \in \M, \text{ otherwise} \end{cases}$
\STATE \textbf{OUTPUT:} $\tilde{p} \in \M$
} 
\end{algorithmic}
\caption{\Protect{} and \Access{} of the proposed concrete realization.  }
\label{algo:concrete}
\end{algorithm}

\section{Security Analysis}
\label{sec:analysis}

This section starts with a proof that the security of the scheme relies purely
on the entropy of the selected attributes (if the underlying cryptographic primitives are secure).
As a consequence, we analyze a dictionary attack based on the entropy of selected social
network data.

\subsection{Proof of Security}
\label{sec:securityA}
In this section, we provide an upper bound on the advantage of an attacker. 
Recall that only established cryptographic primitives, i.e. block cipher, hash function, and secret sharing scheme, are deployed. For each of these concrete realizations are known, which are either secure according to the current state of knowledge, e.g. the block cipher standard AES and the hash function standard SHA-3, or can even be mathematically proven to be secure, e.g. Shamir's secret sharing scheme. Hence, to keep the analysis simple, we assume that these schemes are ideally realized.  More precisely, we assume in the following that (i) the block cipher is a random permutation, (ii) the hash function is a random oracle, and (iii) the secret sharing scheme is information-theoretically secure. Moreover, we make the attacker somewhat stronger by considering for the time effort only the number $q$ of queries it makes to the ideal cipher and/or the random oracle.\footnote{We stress that these assumptions are only made to keep the analysis simple. In fact, the analysis could easily be adapted to the case that the deployed block cipher and hash function are ''only'' computationally secure.}

The main theorem is the following:
\begin{theorem}\label{theo:main}
It holds for any adversary $\adv$ that makes at most $q$ queries to the ideal cipher and/or the random oracle that 
\begin{equation}
\mathrm{Adv}_\adv\leq \frac{3q}{|\mathcal{K}|}+Pr[E_v].
\end{equation}
Here $E_v$ denotes the event that $\adv$ correctly identifies for the given attribute descriptions $(d_1,\ldots,d_\n)$ (see Sec.~\ref{sec:concept}) at least $\threshold$ attribute values. 
\end{theorem} 
\paragraph{Proof}
$\adv$ knows the encrypted post $c$, the encrypted shares $c_1,...,c_\n$, and the attribute descriptions $(d_1,\ldots,d_\n)$. 
 Let $E_{mk}$ denote the event that during the attack, the attacker made a query to the block cipher using the correct key. 
In the ideal cipher model, $c$ is equally likely to be an encryption of $p_0$ and $p_1$ if $\neg E_{mk}$. 
Hence 
\begin{eqnarray*}
\mathrm{Adv}_\adv&=&\pr{b=b^*}-1/2\\
&=&\pr{b=b^*\vert E_{mk}}\cdot\pr{E_{mk}} \\
&+& \pr{b=b^*\vert \neg E_{mk}}\cdot\pr{\neg E_{mk}}-1/2\\
&\leq &1\cdot \pr{E_{mk}}+1/2\cdot 1-1/2=\pr{E_{mk}}.
\end{eqnarray*}
Thus, the advantage of $\adv$ is equivalent to the probability $\pr{E_{mk}}$ of deriving the master key $\key$.
Let $s_1,\ldots,s_n$ denote the shares, which encode the key $\key$. Let $E_s$ denote the event that the attacker successfully reconstructed at least $\threshold$ from these $n$ shares. Then it holds
\begin{align*}
\pr{E_{mk}}&=\pr{E_{mk}\vert E_s}\cdot \pr{E_s}+\pr{E_{mk}\vert \neg E_s}\cdot \pr{\neg E_s} \\
&\leq  Pr[E_s]+q/|\K|
%
\end{align*}
The second term follows from the fact that $\pr{E_{mk}\vert \neg E_s}\leq q/|\K|$. The reason is that if the attacker did not recover sufficient shares, it follows from the information-theoretic security of the secret sharing scheme that the attacker cannot derive any information about $\key$, which he did not know before. Thus, the only remaining option in this case is to coincidentally query the block cipher with the correct key. This is successful with probability at most 
$q/|\K|$. 

Next, we consider $Pr[E_s]$. Recall that the attacker only knows the encryption of the shares, being $c_i=\enc(H(v_i),s_i)$ for the selected attribute values $v_i$. Let $h_i:=H(v_i)$ denote their hash values. We denote by $E_h$ the event that the attacker correctly identified at least $t$ hash values $h_i$. It follows that 
\begin{eqnarray*}
\pr{E_s}&=&\pr{E_s\vert E_h}\cdot \pr{E_h}+\pr{E_s\vert \neg E_h}\cdot \pr{\neg E_h} \\
&\leq & 1\cdot \pr{E_h}+q/|\K|.
\end{eqnarray*}
The rationale behind the second term is similar to above: if an attacker does not know all keys that have been used to encrypt the shares, the best he can do is to guess the remaining ones.

Finally, an upper bound for $Pr[E_h]$ is derived. Let $E_v$ denote the event that the attacker identified at least $t$ values $v_i$ correctly. Using the same line of arguments, it follows that 
\begin{equation*}
\pr{E_h}\leq \pr{E_v}+q/|\K|.
\end{equation*}
Assembling all results yields the claim.\qed
%
%

Theorem \ref{theo:main} shows that if a reasonable key size, e.g., 128 bits, are chosen, the advantage of an attacker is dominated by the probability of an attacker of correctly guessing sufficiently many attribute values. Consequently, we consider in the remainder of this section a dictionary attack for guessing these values based on real-world data and distributions for the evaluation.

\newpage
\subsection{Dictionary Attack Evaluation}
\label{sec:securityC}


Having proven that the security depends solely on the probability of determining the chosen values, dictionary attacks that aim to guess those values are the foremost vulnerability for PKA. 
We consequently analyze the feasibility of a dictionary attack based on real-world data. 
Note that this analysis is only valid for automated attacks, most commonly executed by crawlers.
The scheme is vulnerable to social engineering attacks, which invest time in retrieving information about
the person from out-of-band sources. However, such attacks have a high cost, and are not scalable.
Recall that an attacker cannot tell from the fact that he chooses incorrect values, which values
have been correct, if any, and which incorrect. 
Consequently, an unsuccessful attack without additional information about the publisher reveals \schange{very little information about the profile owner.}
{only that at least $\n -\threshold$ values are incorrect. 
The information gain is negligible, due to the vast number of possible
attribute-value combinations.
}
%
The dictionary attack works as follows:
an attacker $A$ uses a \emph{basis distribution} $\distr_B$ of attribute values and aims to guess attribute values which have been sampled according to a (possibly unknown)  \emph{target distribution} $\distr_T$. 
Based on $\distr_B$, $A$ then tries all possible sets of $\threshold$ values
in descending order of likelihood. For each guess, a corresponding key candidate $\tilde{\key}$ is computed and the encrypted post decrypted using $\tilde{\key}$.
We assume that $A$ can distinguish correct and incorrect decryptions, since it is easy to recognize real language from randomly generated character strings. 
Hence, $A$ continues testing different sets of values until the encrypted post is correctly decrypted\schange{or a maximal number $c$ of trials is reached}{}.
In the following, we refer by \Aplain{\distr_B}{t}{\distr_T} to an attacker on a  $t$-of-$4$ scheme with target distribution $\distr_T$, using basis distribution $\distr_B$. 
Two metrics are important for measuring $A$'s efficiency: the fraction of protected posts that can be accessed using these predefined values, and the cost, meaning the number of trials, $A$ has to invest to find the correct values.

\paragraph{Datasets and Distributions}
We use two available datasets to analyze the efficiency of possible attacks:
the first was obtained from meinVZ, a German OSN with strong similarities to Facebook \cite{Bilge2009all}, the other from the German business-oriented social network XING \cite{strufe10popularity}.

Both datasets contain the attributes descriptions 'first name', 'second name', 'home town', 'ZIP' and 'university'. 
We chose the $266,804$ profiles that contain all considered attributes from the entire set of $702.986$ profiles in meinVZ, and analogously  $42,475$  profiles of the overall $756,400$ profiles in XING respectively.
Since 'hometown' and 'ZIP' are redundant, we use 'hometown' only, especially as the majority of users decided not to enter their ZIP in XING. 
To increase the attacker's chances, we preprocessed the data sets by by adding variants of attribute values,  for example 'Uni' instead of 'university'.
We additionally obtain the overall German statistics as published by the German statistical office \footnote{\fnsize{\url{https://www.destatis.de/DE/Startseite.html}}} (henceforth denoted {\em SB}).

For our analysis, we assume that each profile contains a post protected with these four respective attributes and that these attributes are independent.
Note that information on hometown and university may be publicly available and hence unsuitable choices for the
attributes in reality, so they are just used for the purpose of our evaluation for the lack of other data.
However, names, places, and institutions are frequently associated with a common past history 
(name of the favorite teacher, the destination of a school trip, ...), so the respective distributions provide a meaningful analysis of at least a common subset of attribute types.

In the following, $\distr_T(\mathrm{meinVZ})$ and $\distr_T(\mathrm{XING})$ denote target distributions, and $\distr_B(\mathrm{meinVZ})$ and $\distr_B(\mathrm{XING})$ denote the
basic distributions obtained from meinVZ and XING by considering each attribute distribution independently.
For the basis distribution, we calculate the frequency distributions of the attribute values for  meinVZ, XING and SB, respectively. Table \ref{tab:dist} displays the number of different attribute values for each data set.
\begin{table}
\centering 
\small
\begin{tabular}{l||l|l|l|l}
Basic Distribution & $1^{st}$ Name & $2^{nd}$ Name & Uni & Town \\
\hline
SB & 1878 & 3422 & 568 & 688 \\
meinVZ & 23139 & 98013 & 2861 & 52989 \\
XING & 4590& 24305 & 16423 & 4686 \\
\end{tabular}
\caption{Number of values for each dataset}
\label{tab:dist}
\end{table} 

The order of likelihood is evaluated as follows: For a fixed attribute description, a possible attribute value has  
 rank $r$ with regard to the basis distribution $\distr_B$ if it is the $r$-th most likely value according to distribution $\distr_B$.
Hence, for  given attribute descriptions $d=(d_1,\ldots,d_\threshold)$, the rank of a possible assignment of attribute values $v=(v_1,\ldots ,v_\threshold)$ is estimated as the product of the ranks of the individual $v_i$'s, that is $rank(v)$ is estimated by $\prod_{i=1}^n rank(v_i)$.
Note that this is a lower bound.
We compute estimates only, since it is computationally demanding to compute and store a distribution over several billions
of values. 


\paragraph{Success rate}
\begin{table}
\begin{tabular}{l|l|r|r|r}
 & & \multicolumn{3}{|c}{Basic Distribution} \\
Target Dist. & $\threshold$ & SB & meinVZ & XING  \\
\hline
\multirow{4}{*}{meinVZ} & 1 & 0.6390 & 1.0 & 0.9714 \\ 
& 2 & 0.1923 & 1.0 &0.8172 \\ 
& 3 & 0.0162& 1.0 & 0.4804\\ 
& 4 & 0.0 & 1.0 & 0.1339 \\ 
\hline
\multirow{4}{*}{XING} & 1 & 0.7968 & 0.9973 & 1.0 \\ 
& 2 & 0.3612 & 0.9604 &1.0 \\ 
& 3 & 0.0543 & 0.6880 & 1.0\\ 
& 4 & 0.0 & 0.0746 & 1.0\\  
\end{tabular}
      \captionof{table}{Fraction of accessed posts for different distributions}
      \label{tab:successrate}
\end{table}
For a set of posts, a basis distribution $\distr_B$ and a threshold $\threshold$, the \emph{success rate} is given by the fraction
of posts that can be accessed.
The fractions of accessible posts for each set of posts and each distribution are given in Table \ref{tab:successrate}. 
Naturally, an attacker of the form \A{OSN}{t}{OSN} can access all posts as the basis distribution contains all attribute values occurring in the target distribution.
Moreover, as expected the success rate is the lowest if using $\distr_B(\mathrm{SB})$ as $\distr_B(\mathrm{SB})$ contains less values than the other basis distributions.
Even with a threshold of $\threshold=1$, \A{SB}{1}{meinVZ} only has a success rate of $64$ \%, whereas \A{SB}{1}{XING}
has a success rate of about $80$ \%.
This means that a considerable fraction of the users did not enter a single value that exists in official statistics. 
The success rate decreases rapidly with increasing $\threshold$, so that for $\threshold=4$, no post can be accessed.

Using a basic distribution crawled from a different OSN, more than $80$\% of the posts can be accessed using a $1$-of-$4$ or $2$-of-$4$ secret sharing scheme.
 However, when requiring all $4$ attributes, we have a success rate of $7.46$ \% for \A{XING}{4}{meinVZ} and of 
   $13.39$ \% for \A{meinVZ}{4}{XING}, respectively.
So, as expected, for both SB and OSN basic distributions, the success rate decreases enormously when increasing $\threshold$. Though a considerable number of posts can be accessed with a threshold of 1 or 2, less than $15$ \% of the posts can be accessed with $\threshold \geq 3$.

\begin{figure}
    \centering
    \includegraphics[width=0.95\linewidth]{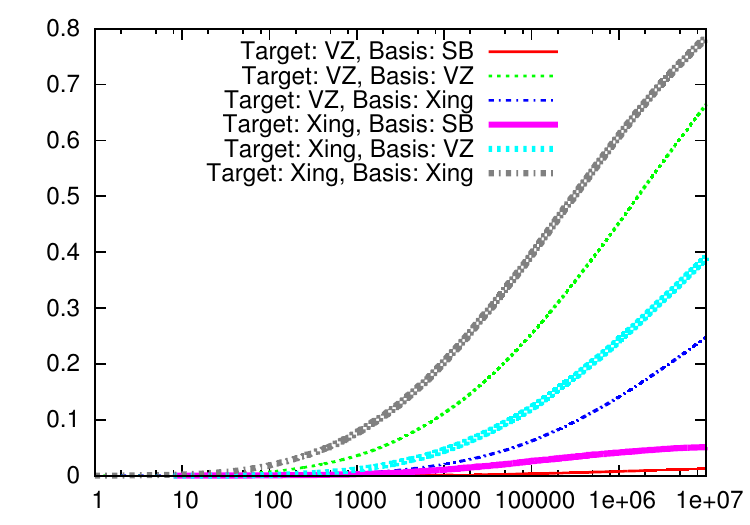}
    \captionof{figure}{Number of needed trials vs. fraction of accessed posts for a $3$-of-$4$ scheme}
    \label{fig:3-4}
  \end{figure}
\paragraph{Mean Number of Trials}
The mean cost of an attack is given as the estimated number of trials, averaged over all posts, 
with regard to a basic distribution $\distr_B$ and a threshold $t$.
Table \ref{tab:cost} details the average cost together with the standard deviation. 
In case of $\distr_B(\mathrm{SB})$, the cost for $\threshold=1\ldots 3$ lies on the order
of at least $10^{11}$ while the cost for $4$-of-$4$ equals the number of possible combinations (about $2*10^{12}$), because no post can be accessed. 
Nevertheless, this is lower than the cost for \A{meinVZ}{t}{XING} and \A{XING}{t}{meinVZ} for $\threshold=1\ldots 4$.
The reason for this lies again in the lower number of values considered by the SB basic distribution.
In both cases, the cost is at least on the order of $10^{14}$, even for $\threshold=1$. The cost when using XING as a basic distribution and
meinVZ as a target distribution is lower than in the opposite case, due to the fact that $\distr_B(\mathrm{SB})$ considers less values. 
In case of an attack of the form \A{OSN}{t}{OSN}, the cost is considerably reduced (see Table \ref{tab:cost}).

%


\begin{table*}[ht]
\small
\begin{center}
\begin{tabular}{l|l|r|r|r}
 & & \multicolumn{3}{|c}{Basic Distribution} \\
Target Dist. & $\threshold$ & SB & meinVZ & XING \\
\hline
\multirow{4}{*}{meinVZ} & 1 &  9.06E11 $\pm$ 1.21E12 & 6.53E01 $\pm$ 1.52E02 & 2.45E14 $\pm$ 1.43E15 \\ 
& 2 & 2.03E12 $\pm$ 9.90E11 & 1.14E05 $\pm$ 9.73E05 & 1.57E15 $\pm$ 3.32E15 \\ 
& 3 & 2.47E12 $\pm$ 3.17E11 & 2.24E09 $\pm$ 2.98E10 & 4.46E15 $\pm$ 4.29E15 \\ 
& 4 & 2.51E12 $\pm$ 0.00E00 & 1.30E14 $\pm$ 1.93E15 & 7.44E15 $\pm$ 2.92E15 \\ 
\hline
\multirow{4}{*}{XING} & 1 & 5.10E11 $\pm$ 1.01E12 & 9.15E14 $\pm$ 1.77E16 & 4.98E01 $\pm$ 2.01E02\\ 
& 2 & 1.60E12 $\pm$ 1.21E12 & 1.36E16 $\pm$ 6.70E16 & 8.27E04 $\pm$ 6.75E05\\ 
& 3 & 2.37E12 $\pm$ 5.69E11 & 1.07E17 $\pm$ 1.59E17 & 5.84E08 $\pm$ 6.13E09\\ 
& 4 & 2.51E12 $\pm$ 0.00E00 & 3.18E17 $\pm$ 9.03E16 & 9.13E12 $\pm$ 1.06E14 \\ 
\end{tabular}
\caption{Mean number of trials needed to access a post using a dictionary attack for three distinct value distributions (E notation)}
\label{tab:cost}
\end{center} 
\end{table*}

\paragraph{Payoff Success vs. Invested Time}
The mean cost is only of limited value, especially since the standard deviation is high. 
An adversary is more interested in minimizing the effort for maximum success. 
Hence, $A$ will only try a certain number of values before aborting and trying a different post.
This allows us to compare different distributions with regard to their efficiency. 
Figure \ref{fig:3-4} displays the effort in terms of needed of trials versus the likelihood of accessing a post using up to $10^7$ trials (for higher values the increase is barely noticeable). 
Indeed, the curves have a strong increase in the beginning, as expected. 
Using the same basic and target distribution is more efficient than using data from a different OSN.
However, the latter still produces a higher success rate for any number of trials than the SB distribution. 
This is due to the frequent use of inofficial names in profiles, such as storing the nickname as a users first name. 
Figure \ref{fig:3-4} indicates that XING posts can be accessed within a lower number of trials, at least for $\threshold=3$. 
A plausible reason is that XING is addressed to professionals, meaning users in general use less inofficial names, so having more common attribute values and a smaller set of values.


\paragraph{Summary} For a threshold $\threshold=3$, at least $100$ million trials are needed in average to access a post. Though this can be realized, it is not within the computational power of a spammer or an advertisement company, providing a good security against this kind of adversary. Moreover, the entropy of our scheme seems to be similar to that of actual selected passwords \cite{malone:password,Narayanan05fast}.

Recall that our analysis assumes that values with a low entropy are chosen. As proven in Th.~\ref{theo:main}, the security depends solely on the entropy of the selected attribute values.
That is already a $1$-of-$1$ scheme with an arbitrary character string as value is
 secure, meaning it is not possible to determine the protected content. 
In this fashion, our scheme could be used in the same way as traditional encryption, providing the same guarantees as
any well-established cryptographic primitives.
At the end, it is up to the concrete scenario to choose a balance between security and probability that a legitimate user can determine the requested attribute values. 


\section{Implementation}
\label{sec:plugin}

To give proof of our concept, we implemented \KDAF{}, em\-ploy\-ing Shamir's secret sharing \cite{Shamir1979SS}, AES-256, and the SHA-256 hash function, as a Firefox extension in JavaScript.
It is available for download
\footnote{\fnsize{{\url{http://www.p2p.tu-darmstadt.de/research/PKA/}}}} 
and it can be used to publish protected posts in Facebook notes, through a simple graphical user interface.
We evaluate the performance of our implementation of \KDAF \ with regard to the overhead for key generation, encryption and decryption. The evaluation shows that the overhead is negligible for an informed user, but prohibitive for an adversary.

\subsection{Extension Description}
\label{sec:pluginDesc}

The extension directly implements the \Protect{} and \Access{} functions. More precisely, the interface provides input methods to share and retrieve protected posts, including the definition of the parameters $\n$ and $\threshold$.
It further allows for addition, modification, and deletion of attribute-value pairs, which are stored locally.
The interface of the extension is defined using XUL, and all algorithms are implemented in JavaScript, relying on existing, verified libraries.
All message exchange leverages the Facebook developer API, and the messages are encoded as JSON objects.

\begin{figure}
\centering
\includegraphics[width=0.8\linewidth]{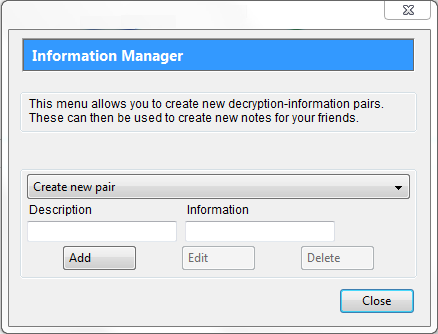}
\caption{Dialog for managing attributes}
\label{fig:manager}
\end{figure}

\begin{figure}
\centering
\includegraphics[width=0.95\linewidth]{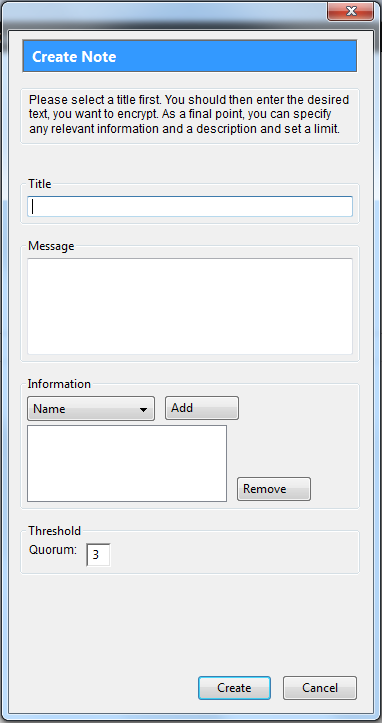}
\caption{Dialog for protecting a note}
\label{fig:encrypt}
\end{figure}

Upon input of a note for protection as well as the corresponding attribute-value pairs and parameters, 
the extension executes \Protect{}.
Attribute-value pairs can be defined in the interface (see Fig.~\ref{fig:manager}).
In order to publish a protected post, several attribute-value pairs can be chosen, and the cleartext of the post is entered (see Fig.~\ref{fig:encrypt}).

The extension then encrypts the post and publishes the ciphertext within Facebook as a note, together with the encrypted shares and the attribute information.
A PRNG provided by the Gibson Research Corporation\footnote{\fnsize{\url{https://www.grc.com/otg/uheprng.htm}}} is used for generating a 256 bit key $\key$. Encryption is done using the AES-256 implementation of the CryptoJS library\footnote{\fnsize{\url{http://groups.google.com/group/crypto-js/topics}}} in CBC mode, and the SHA-256 implementation from the same library is used for hashing.
\key{} is split applying Shamir's secret sharing scheme, which leverages the fact that a polynomial of degree $\threshold-1$ is uniquely identified by at least, but not less than $\threshold$ evaluations. We implement the finite field \gf{} within our extension, and the
secret is split into packets of 8 bits, each representing an element in \gf{}.
For each of these packets, shares are computed and their concatenation then is encrypted using AES.

\begin{figure}
\centering
\includegraphics[width=0.8\linewidth]{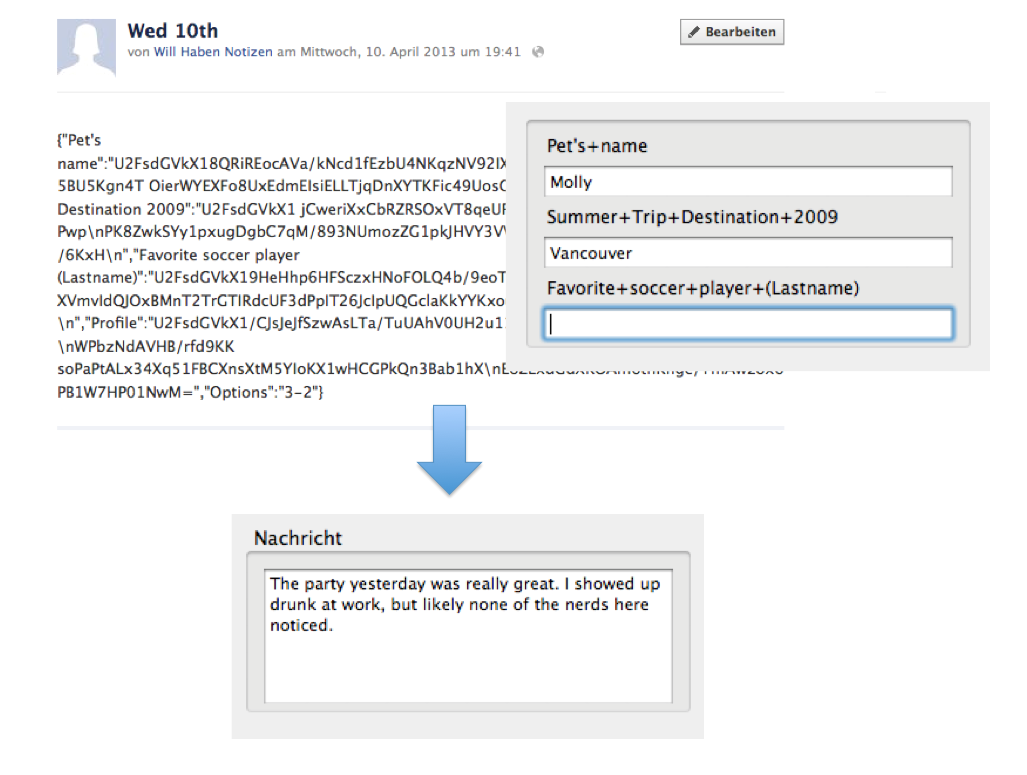}
\caption{A protected note, the attributes, and the accessed content displayed by the extension}
\label{fig:post}
\end{figure}

\begin{figure}
\includegraphics[width=0.95\linewidth]{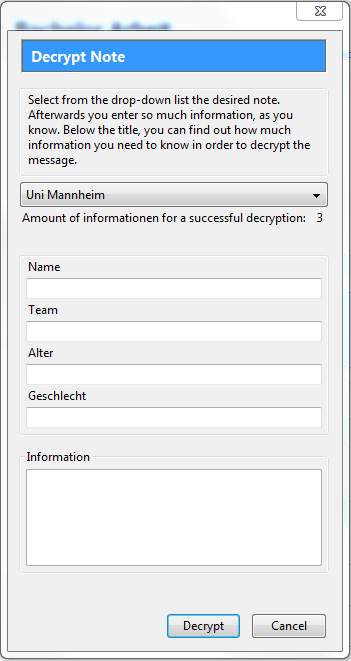}
\caption{Dialog for accessing a note}
\label{fig:decrypt}
\end{figure}

\begin{figure}
\centering
\includegraphics[width=\linewidth]{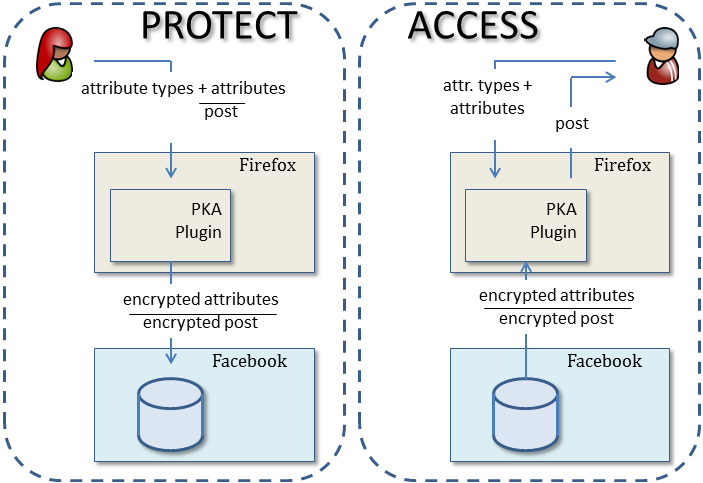}
\caption{Interaction between user, extension and Facebook}
\label{fig:plugin}
\end{figure}

Upon access of a protected post, the user is presented with the requested attributes and can enter the values in corresponding text fields (see Fig.~\ref{fig:decrypt}).
A key is then constructed, and the post decrypted. In case the input values are entered correctly, the original post is displayed, otherwise the AES decryption yields the UTF-8 representation of the attempted decryption (cmp. Fig. \ref{fig:plugin} resp. Fig.~\ref{fig:post}).
Attribute-value pairs both of the user and previously accessed protected posts are stored on the machine of the user only, using Mozilla's SQLite API.



\subsection{Performance Evaluation}
\label{sec:overhead}
In order to test the applicability of our extension, we evaluate its performance on key generation as well as en- and decryption.
We subsequently measure the time needed to access protected posts both for informed users and uninformed attackers, based on the analysis in Section \ref{sec:analysis}.

\begin{figure*}[t]
\centering
\subfloat[Key generation]{\label{fig:keygen}\includegraphics[width=0.33\linewidth]{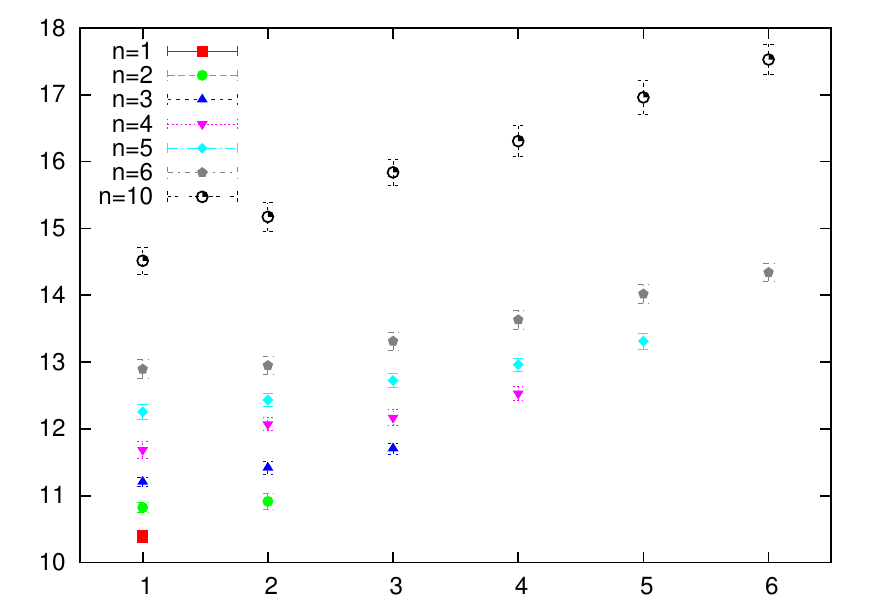}}
\subfloat[En-/Decryption]{\label{fig:encdec}\includegraphics[width=0.33\linewidth]{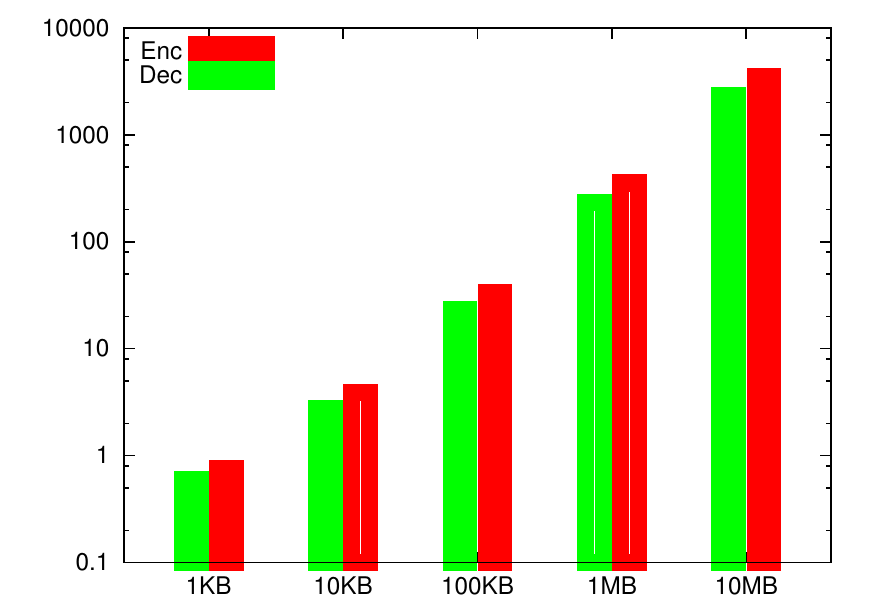}} 
\subfloat[Total Time for Access]{\label{fig:total}\includegraphics[width=0.33\linewidth]{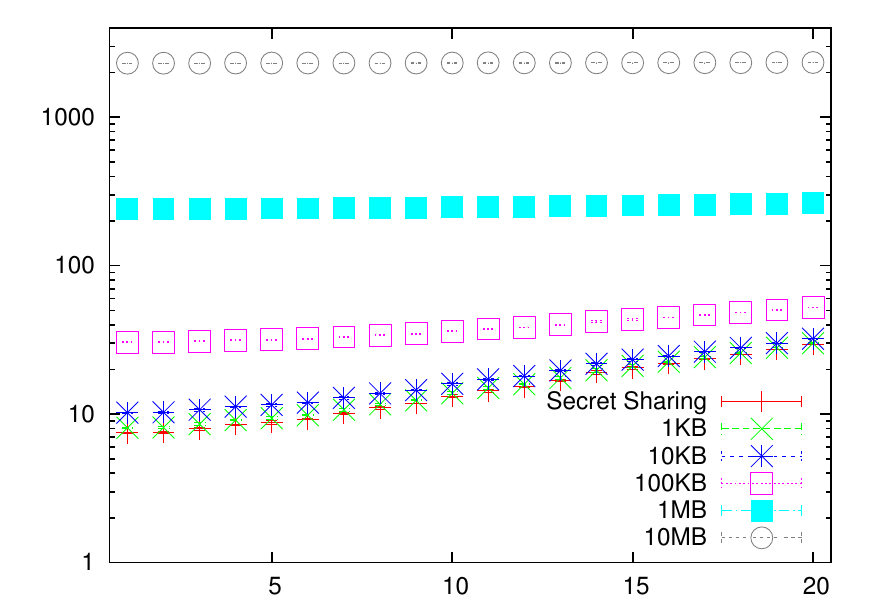}}
\caption{Time in ms needed for a) share generation and encryption for various $\threshold$-of-$\n$ schemes, 
          b) AES en-/decryption with given \key{} for posts between $1$ KB and $10$ MB (log-log), c) total time for access for $\n=20$ and various thresholds $\threshold$ (log-log)}
\end{figure*}

Our measurements are performed on an average dual core PC with 3.00 GHz, 3.50 GB main memory and Windows 7 64 bit as operating system.
Results are averaged over $100$ runs and presented including their standard deviation.
The duration of cryptographic operations depends on the total number of attributes $\n$,
the number of attributes needed to access the post $\threshold$, as well as the post size.

We vary the number of shares $\n$ between $1$ and $20$, whereas $\threshold$ is chosen between $1$ and $\n$.
The following elementary operations are evaluated: 
\begin{enumerate}
 \item generation of keys (including share generation and encryption)
 \item encryption and decryption of a post with known \key
 \item reconstruction of \key{} from the encrypted shares.
\end{enumerate}
The total time for access is the sum of key reconstruction and the decryption.
We evaluate the en-/decryption cost for posts of sizes $1$ KB to $10$ MB, where the user input is chosen as random UTF-8 character strings.

\paragraph{Key generation}
The key generation consists of the generation of the master key $\key$, the generation of the ($\threshold-1$)-degree polynomial $p$,
the calculation of $\n$ shares and their encryption.
Generating the master key takes $11.61$ ms on average, with a standard deviation of $0.17$ ms.
Fig. \ref{fig:keygen} displays the measured times needed for share generation and encryption
for $\n \in \{1,2,3,4,5,6,10\}$ and $\threshold \in \{1,2,3,4,5,6\}$.

The time for generation and encryption indeed increases linearly with the number of attributes and the number of shares, but it remains below $15$ ms for $\threshold=1$ in all experiments.
Increasing the threshold to $5$ shares, the time ranges from $13.67$ ms ($\n=5$) to $16.47$ ms ($\n=10$). 
In case of 20 shares, the average time for a $5$-of-$20$ scheme is $25.78$ ms, but increases
to $42.86$ ms when a threshold of 20 is used.
In general, key generation and share encryption takes less than $45$ ms, even in case of a scheme with $20$
shares, which seems unlikely high for realistic cases. 

The key generation, secret sharing, and share encryption hence takes too little time to even be perceived by the user.


\paragraph{En-/Decryption with known $\key$}

We measure the duration of en-/decryption for posts sizes increasing by a factor 10 in each step.
Fig. \ref{fig:encdec} indicates that the duration indeed increases linearly. 
The increase from $1$ KB to $10$ KB is slightly lower than in the other cases.
The reason for this deviation is the fact that initial function call and processing the results demand a constant overhead independent of the size of the post.
In case of pure textual posts the size is unlikely to exceed $100$ KB\footnote{\fnsize{{Note that Facebook notes are limited to a length of 65535 characters, yielding a maximum size of $128$ KB UTF-8 encoded text.}}}, in which case the encryption takes well below $35$ ms.
Even in case of a hypothetical $10$ MB post, including several photos for example, the encryption takes only slightly over 3s, which is on the order of time that downloading the same amount of data takes as well.
Note that the median latency of accessing a webpage by a Desktop browser
has been found to be $65$ ms in 2012 \footnote{\fnsize{\url{http://www.webperformancetoday.com/2012/04/02/mobile-versus-desktop-latency/}}}, which is higher than the sum of key generation and encryption cost for realistic post sizes.




\paragraph{Key reconstruction}
Reconstructing \key{} is expected to depend on $\threshold$ for decrypting the shares and
performing the interpolation. 

Fig. \ref{fig:total} displays the total time needed for accessing a post at various post sizes. 
These include the time for key reconstruction and decryption, and the former can be derived by subtracting the decryption time as measured above.
The master key thus is reconstructed in less than $8$ ms at $\threshold \leq 3$, which we argue to be realistic in practice. 
The time for key reconstruction remains below $20$ ms up to a threshold of $14$, and even for $\threshold = 20$ only roughly $30$ ms are needed.



\paragraph{Total Time to Access}
For \KDAF{} to be accepted, the total time for accessing a post is most critical.
This is due to the facts that
\begin{enumerate}
 \item \Protect{} is only executed once per post, while \Access{} is executed by everyone interested in the post, and
 \item an attacker accessing a protected post has to expect a delay of the number of needed trials times the average time for decryption. 
\end{enumerate}
Hence, accessing a post should be fast enough not to decrease the quality of service for a legitimate user, but make it unprofitable for an attacker to attempt gaining access.

Fig. \ref{fig:total} displays the total time access, consisting of key reconstruction and decryption,
for various post sizes. 
It indicates that for small post sizes, the time for share retrieval increases noticeable with the threshold. 
A threshold of $\threshold=20$ takes around 10 times as long as threshold $\threshold=1$ (for a post size of $1$ KB). 
The access time, however, generally remains on the order of hundreds of ms, and the delay of regular access is not noticeable.
In case of large posts exceeding $1$ MB, the delay of the key reconstruction is negligibly in comparison with
the decryption, regardless of the secret sharing parameters.

Legitimate users may need several attempts on the first try until the correct values have been entered, of course. 
This is inevitable since a lot of values are bound to have synonyms. 
Assuming that such synonyms are few and that users carefully choose their values, we expect that a maximum of 10 combinations will have to be tested.
With content amounting to sizes of several kilobytes, the total access time is still on the order of one second.
The actual perceived overhead by the user hence clearly will be dominated by the time to manually select values for attributes, rather than the time needed by the processing of the extension. 
However, this cost is only applicable for the first trial of a user's keys. Assuming that attributes are frequently reused for protecting posts, all later posts are automatically decrypted by the extension, so that the latency is only that of the cryptographic operations, which is $31$ ms for a 3-of-4 scheme and a post of $100$ KB.  

An uninformed attacker needs to consecutively guess attribute values, decrypt the shares, reconstruct candidates for \key{}, and apply the decryption algorithm.
Simplifying the estimation, we consider only the time needed to retrieve the shares, since the decryption depends on the size of the post.
The results above show that in a $3$-of-$4$ scheme, which seems realistic, a trial to retrieve the shares takes about $7.98$ ms in average.
Consulting the results from Section \ref{sec:securityC}, we can estimate the time the attacker needs to successfully gain access to a post.
%
Using the data sets meinVZ and XING, as described in Section 
\ref{sec:securityC}, we consider value distributions derived
from both OSNs as well as from the German statistical office (SB).
Note that on average at least 54 days are needed to successfully access
a protected post, for all considered basic and target distributions pairs.
Since an attacker is not likely to test for all possible attributes, let us assume that he will try likely combinations for a minute, before turning to the next post.
We see that, even in the case that the ground truth is known, the attacker
is incapable of accessing about 90 \% of the posts within one minute.
The actual distribution, however, is not known to the attacker in reality. 
The best chance of the attacker is to assume that the distributions comply to corresponding official statistics.
Considering the frequency distribution as found in the dataset of the German statistical office in more detail, however, only about $0.08 \%$ of the meinVZ and
less than $1 \%$ of the XING profiles can be accessed in one minute.


The performance analysis shows that the protection and accessing time is
barely noticeable for a legitimate user, but, assuming a sensible
attribute selection, an attacker needs to invest days into 
accessing single posts. This represents a burden none of our considered adversaries - crawlers, third party applications,
or the OSN provider - are willing to invest.



\section{Conclusion}
\label{sec:conclusion}

We have introduced PKA, a non-interactive scheme to protect public OSN data from crawlers\schange{, third-party applications, OSN providers, and curious strangers. 
PKA combines protection against surveillance by global adversaries with protection against accidental oversharing.}{.
Though it is not the main application, protection against third-party applications, OSN providers, and arbitrary individuals can be achieved.} 
User implicitly select their intended audience and level of security by their choice of attributes and values.
Rather than replacing common access rules, PKA is an extension to the traditional layers of friends, friends-of-friends, and strangers, offering an additional, more fine-grained
grouping of contacts. 
We have shown that PKA is  secure under the assumption that the underlying cryptographic primitives - secret sharing scheme, symmetric encryption, and hashing -
are secure.
Furthermore, we estimated the number of trials needed to decrypt one post on the basis of data crawled from two OSNs. 
Even with rather unsuitable attribute choices, access is delayed on the order of minutes to days.
Our Firefox extension, implementing PKA as proof of concept, supports the applicability of PKA, exhibiting negligible delays for legitimate access.

However, the current implementation focuses on providing the basic functionalities at a high
level of security rather than usability. 
We plan on extending its functionalities to enable the encryption of non-textual content such 
as pictures.
In addition, it should be easily possible to encrypt posts with complete previously used settings,
whereas currently it is only possible to reuse attribute-value pairs.
From a conceptional perspective, we currently evaluate how to use general access secret sharing 
 \cite{itoh87secret} for weighted attributes.

\bibliographystyle{unsrt}
\bibliography{main}
%
\end{document}